\documentclass[
    ,final            % use final for the camera ready runs
    ,numberedheadings % uncomment this option for numbered sections
  ]
  {aipproc}

\layoutstyle{6x9}

\begin{document}
\newcommand{\apj}{Astrophys. J.}
\title{Neutralino annihilation in the Large
Magellanic Cloud}

\author{Argyro Tasitsiomi}{
  address={Kavli Institute for Cosmological Physics and Department of Astronomy \& Astrophysics\\
  University of Chicago, 5640 S. Ellis Avenue, Chicago, IL 60637, USA}
}

%\author{<author2>}{
%  address={<common address for author2 and author3>}
%}

%\author{<author3>}{
%  address={<common address for author2 and author3>}
%  ,altaddress={<author1 address>} % additional visiting address
%}

\begin{abstract}
I present results on the expected  $\gamma$-ray and synchrotron fluxes  due to
neutralino annihilation in the Large Magellanic Cloud based on 
\cite{tasitsiomi_etal04}.
%The dark matter density profile of the LMC is obtained by fitting its rotation velocity
%curve using models predicted by N-body simulations and allowing for tidal stripping effects.  
%The $\gamma$-ray 
%flux expected from these models may be detectable by GLAST for a significant part 
%of the neutralino parameter space.  The prospects 
%for existing and upcoming Atmospheric Cherenkov Telescopes (ACTs) are less optimistic, 
%as unrealistically long exposures are required for detection. 
%The maximum possible flux predicted is well below EGRET's measurements and thus  EGRET does not constrain the parameter space.  
%The expected  synchrotron emission generally lies 
%below the observed radio emission from the LMC in the frequency range of 19.7 to 8550 MHz.  
%The low level of both the $\gamma$-ray and synchrotron emission strengthens the usual assumption that most
%of the observed signal is due to cosmic rays. However, the different radial and energy dependence of neutralino
%and cosmic ray induced signals, as well as the different radii of dominance for each of these signals,
%are to be promising  in help disentangle the neutralino signal from the cosmic ray induced one. 
\end{abstract}

\maketitle

%%%%%%%%%%%%%%%%%%%%%%%%%%%%%%%%%%%%%%%%%%%%
%% MAINMATTER
%%%%%%%%%%%%%%%%%%%%%%%%%%%%%%%%%%%%%%%%%%%%
\section{Introduction}
Decades of observational evidence indicate that galaxies
are surrounded by massive dark matter halos.  
Of the many candidate dark matter particles that have been proposed, the most popular is the neutralino, $\chi$. 
Products of $\chi\bar{\chi}$ annihilation rapidly
decay into neutrinos and $\gamma$-rays, as well as electrons and positrons which emit synchrotron radiation in local magnetic fields.
At a distance of 50.1 kpc~\cite{vandermarel_etal02}, and given a wealth of observations indicating the existence
of significant amounts of dark matter \cite[e.g.,][]{vandermarel_etal02,sofue_99}, the LMC is an
obvious choice for neutralino detection. 
 
The flux of both $\gamma$-ray and synchrotron emission from neutralino annihilation in the LMC 
depends on the square of the density profile of the LMC dark halo. 
To obtain the density profile we fit the rotation velocity 
HI data from  Kim et al.~\cite{kim_etal98} and carbon star  data 
from Alves and Nelson~\cite{alves_nelson00}.  
We considered a range of density profiles, but here I discuss only  the Navarro, Frenk and White (NFW)~\cite{navarro_etal95,navarro_etal96}
profile, 
%All three fits to the rotation curve data are shown in Fig.\ \ref{fig:rc}.
%While all the fits are acceptable, the Hayashi et al.\ profiles better fit the data
%at the outer radii.  The masses predicted by these models are in reasonable
%agreement with estimates in the literature -- the Hayashi et al.\ gives
%a mass of  $\sim 9 \times 10^9 M_{\odot}$ within $8.9$ kpc, while the NFW fit
%gives $\sim 3 \times 10^{10} M_{\odot}$ within 8.9 kpc, slightly above the observationally determined $(8.7 \pm 4.3) \times 10^9  M_{\odot}$ within 8.9 kpc by~\cite{vandermarel_etal02}.  
%\begin{figure}
%  \includegraphics[height=.3\textheight]{figure1}
%  \caption{Rotation curve of the LMC: square points from
%HI data~\cite{kim_etal98} and triangle points from carbon star data~\cite{alves_nelson00}. The crossed curve is the NFW fit. 
%The dashed curve is the Hayashi et al.\ fit obtained by fixing the scale radius to the
%best fit value found for the NFW profile.  The solid line corresponds to the Hayashi et al.\ fit obtained by fixing the scale radius at $r_{s} \sim 1.4 r_{s,NFW}$ to 
%account for the possible evolution of $r_{s}$ (see text for details).}
%\label{fig:rc}
%\end{figure}
and a shallower profile -- an isothermal sphere with a core as derived in \cite{alves_nelson00}
by fitting the same LMC rotation curve data we used.
%Both  the Moore et al.~\cite{moore_etal98} and the 
%Stoehr et al.~\cite{stoehr_etal02} profiles 
%seem less favored by the rotation curve data.
This  profile represents a less ideal scenario.  
Instead of making the assumption of a minimum disk, the stellar 
disk and gas contributions are included in the model, hence the contribution from the dark matter is confined to smaller values compared with
the minimum disk assumption.  
More details on the various fits to the observed rotation curve are given in \cite{tasitsiomi_etal04}.
\section{The $\gamma$-ray Emission}
Neutralino annihilation proceeds through a number of channels, many of which
produce $\gamma$-rays in the final state (see e.g., \cite{cesarini_etal03}).  
We considered only the continuum $\gamma$-ray emission due to the decay of neutral pions produced in  hadronic jets.  
Given the uncertainties in the dark matter density distribution, 
we  used the approximate Hill spectrum~\cite{hill_83} to calculate the
number of photons above a certain energy threshold, $N_{\gamma}(E>E_{th})$.  
The simplified spectra derived using this approximation vary from the fit obtained in~\cite{bergstrom_etal01} by no more than 
a factor of a few for neutralino masses from 10 GeV up to a few TeV. 
The $\gamma$-ray emission coefficient $j$ is
\begin{equation}
%\vspace{-0.58cm}
\label{gammas_emission}
j=N_{\gamma}(E \geq E_{th}) \frac{\langle \sigma v \rangle_{\gamma}}{m_{\chi}^2} \rho^2(r)\ ,
%%\vspace{0.015cm}
\end{equation}
The quantity 
$\langle\sigma v\rangle_{\gamma}$ is the thermally averaged cross section times
velocity for annihilation
into $\gamma$-rays, $m_{\chi}$ is the neutralino mass, and
$\rho(r)$ is the density profile of the halo. The specific intensities and fluxes, discussed in what follows, are obtained by appropriate
integration of the emission coefficient (along lines of sight and over volume, respectively). For more details,
see \cite{tasitsiomi_etal04}.

EGRET detected a flux of  $(14.4 \pm 4.7) \times 10^{-8}$  
photons $(E > 100\ \rm{MeV})$ cm$^{-2}$ s$^{-1}$ from  the LMC~\cite{hartman_etal99}.  
Using our fit to the NFW profile and an energy threshold of 100 MeV,
the maximum flux produced by a viable SUSY model is $\simeq 3.3 \times 
10^{-9}$ photons cm$^{-2}$ s$^{-1}$, corresponding to $m_{\chi} 
\simeq 50$ GeV and $\langle \sigma v \rangle_{\gamma}
\simeq 2 \times 10^{-26}$ cm$^3$ s$^{-1}$.  
While being consistent with the 
flux detected by EGRET, even the maximum predicted flux is
almost  two orders of magnitude too low, suggesting that the 
primary source is cosmic rays.  
In fact the cosmic ray induced $\gamma$-rays may be an additional background
component to consider when trying to detect flux from neutralino annihilation. Following a method similar to that presented 
in ~\cite{pavlidou_fields01}, we calculated this cosmic ray induced background.
\begin{figure}
\resizebox{.88\textwidth}{!}
{\hspace{-4cm} 
\includegraphics[height=0.6\textheight]{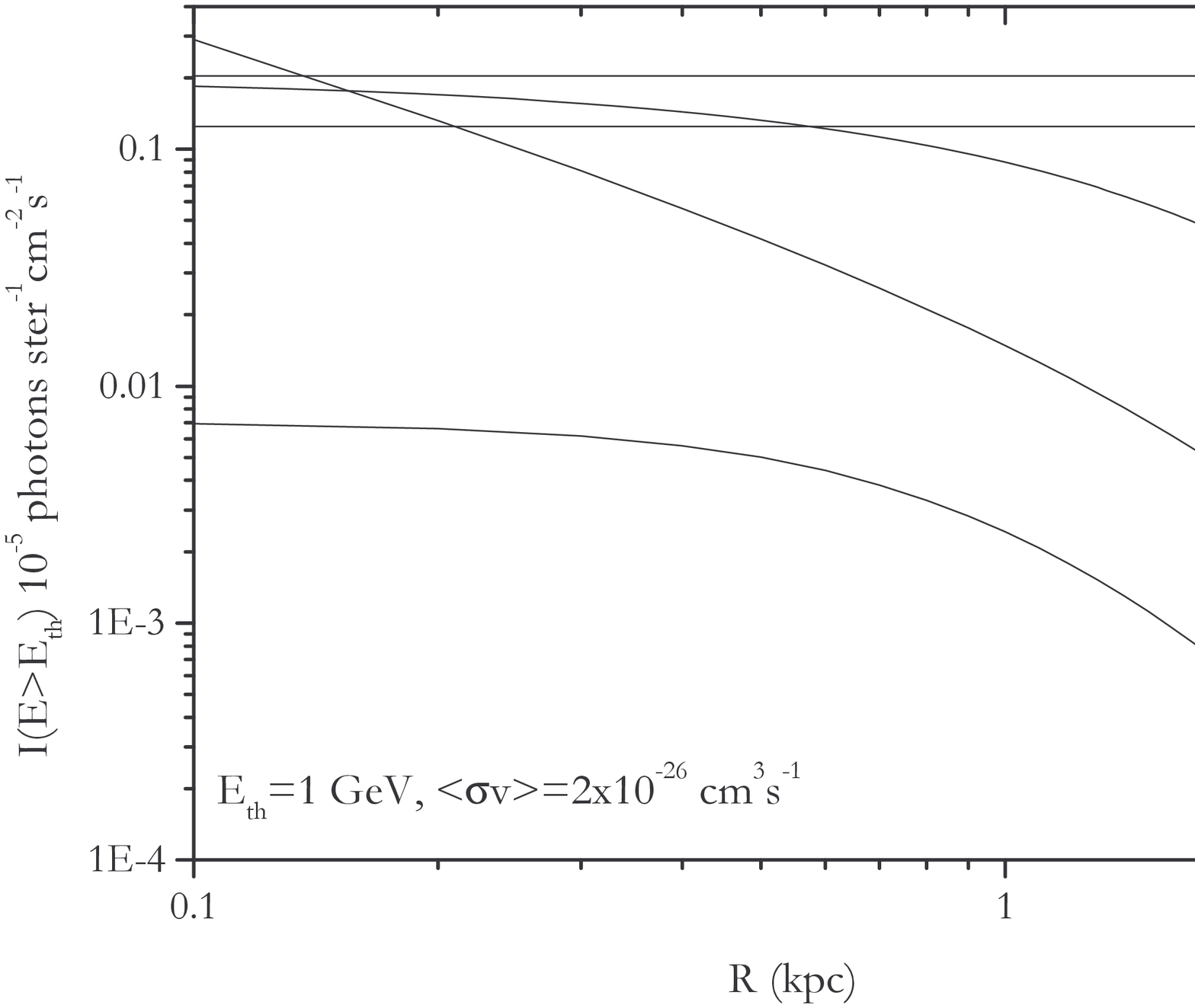}
\hspace{3.5cm} \includegraphics[height=0.6\textheight]{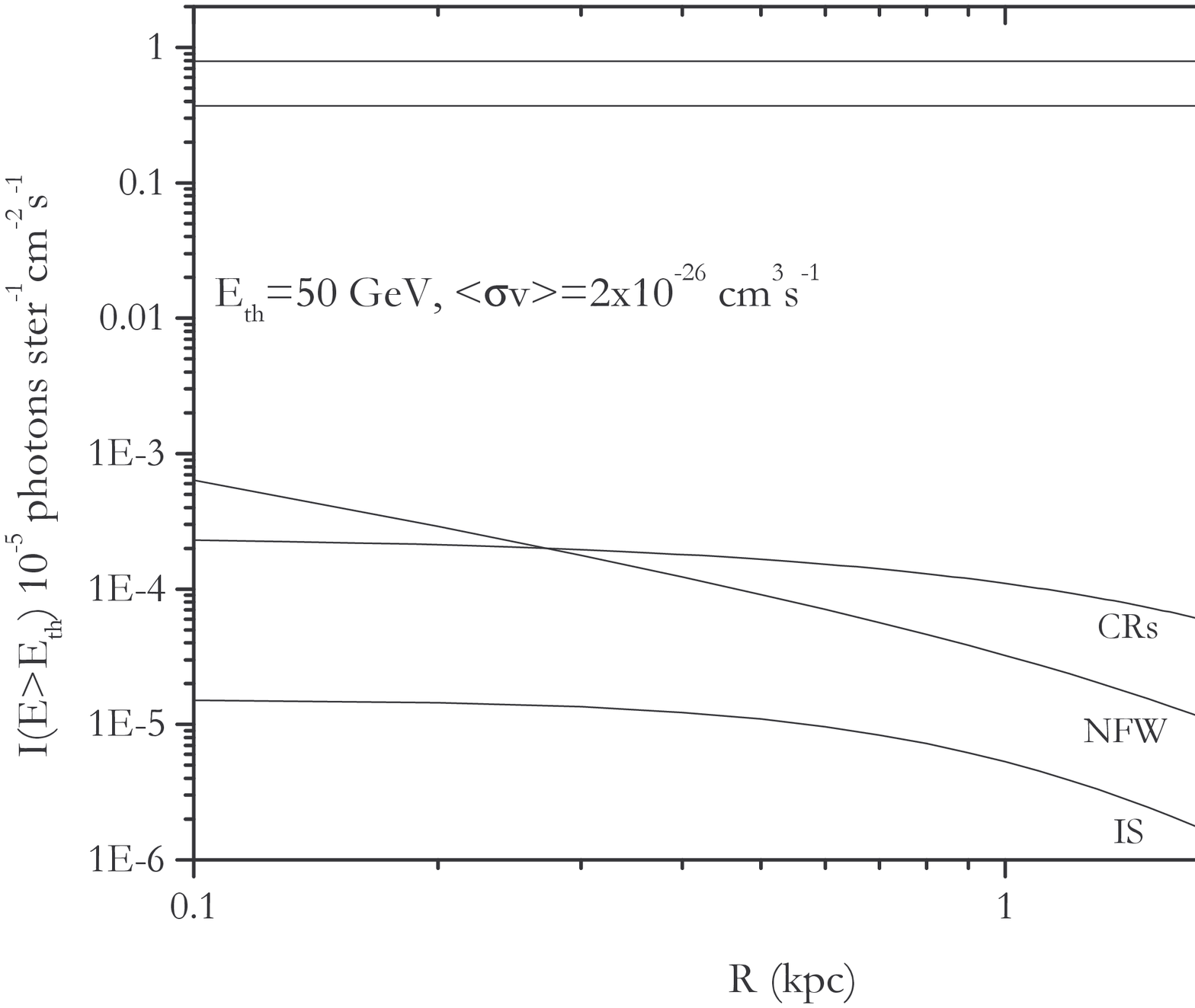}}
%\vspace{-0.3\textheight}  
\caption{
{\it Left panel:} Specific intensity as a function of projected distance from the 
center of the LMC for several background contributions and the neutralino signal above 1 GeV.\
The signal is shown for the NFW profile and the isothermal sphere with a core (IS).
The background components are labeled as follows: G for galactic, EG for extragalactic, CRs for cosmic rays.
{\it Right panel:} Same as left panel but for an energy threshold of 50 GeV. The background components
in this case are labeled as follows: H for hadronic and E for electronic. }
\label{fig:intensity_comp}
\end{figure}

In Figure \ref{fig:intensity_comp} I show a comparison of the specific 
intensity of the different backgrounds, as well as that 
for the NFW profile and the isothermal sphere with a core.
The left panel assumes an energy threshold of 1 GeV which is appropriate for GLAST, 
while the right panel assumes a 50 GeV threshold appropriate for ACTs.
For GLAST  the relevant backgrounds are 
the galactic and extragalactic
diffuse emission~\cite{sreekumar_etal98,bergstrom_etal98} 
. For ACTs the relevant  backgrounds 
are the hadronic and electronic cosmic ray shower 
contributions~\cite{bergstrom_etal98,longair_92}, though it is worth noting that ACTs can reject
hadronic showers with high efficiency.
Clearly, 
the cosmic ray background is not dominant at these energy thresholds. 

\begin{figure}
  \includegraphics[height=.315\textheight]{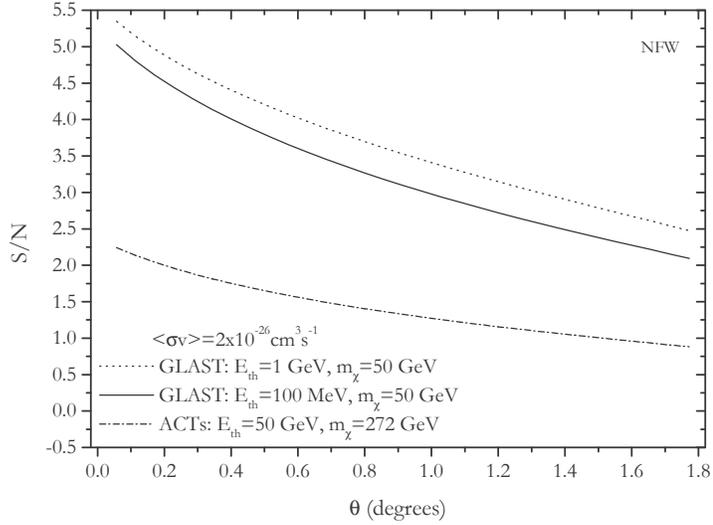}
  \caption{Angular dependence of the signal-to-noise ratio for a NFW dark matter halo profile.
For GLAST we use  $A_{eff}=10^{4}$cm$^{2}$ and 1 year of observation.
For  ACTs we use  $A_{eff}=10^{9}$cm$^{2}$ and 1 month of observation.}
\label{fig:s2n}
\end{figure}

Observationally, the most relevant quantity is the signal-to-noise ratio and its  angular and
energy dependence. In Figure~\ref{fig:s2n} I present 
the signal-to-noise for the flux within an angle $\theta$, assuming the NFW profile.  
Results  are shown for 
three different energy thresholds with the neutralino parameters for each chosen to optimize the signal.
The specifications for each instrument are also indicated.
%this ratio for certain 
%neutralino parameters, chosen to optimize the signal.  
For the noise all the relevant backgrounds, including the cosmic ray
induced emission, were used. From the figure it is clear that focusing on the central regions may help 
to achieve higher
signal-to-noise ratios. As expected, for the isothermal sphere with a core (not shown here),
the results are less optimistic.

\begin{figure}
\includegraphics[height=.315\textheight]{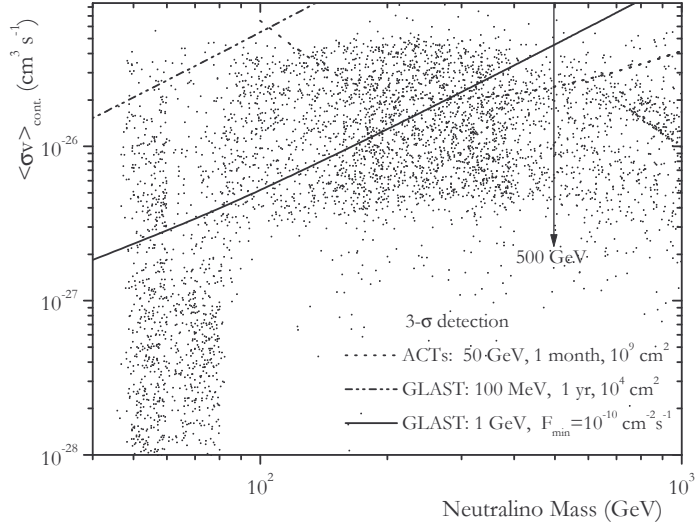}
\caption{The minimum detectable $\langle \sigma v \rangle_{\gamma}$ versus $m_{\chi}$ for the NFW profile. SUSY models above each
curve yield a detectable signal for the instrument and observational 
parameters assumed. The 
dashed line shows detectability by ACTs with an effective area of $10^9$ cm$^2$ 
which will be achievable only at high energy thresholds $(\sim 1$ TeV). The solid line is 
derived for GLAST using the expected flux sensitivity $\sim 10^{-10}$ cm$^{-2}$ s$^{-1}$.  The dot-dashed 
line is derived for GLAST for 1 year of observation, an effective area of $\sim 10^4$ cm$^2$, and
an energy threshold of 100 MeV. }
\label{fig:susy}
\end{figure}

In Fig.\ \ref{fig:susy} I show the parts of the SUSY parameter space that will be detectable. 
The points represent SUSY models produced using the DarkSUSY package~\cite{DarkSUSY}. Models with $m_{\chi}$-$\langle
 \sigma v \rangle_{\gamma}$ above a given line are accessible to observation by the corresponding
instrument, while models in the lower region do not yield a detectable flux. 
For GLAST, the solid line is derived for the optimistic case 
of a flux sensitivity of $\sim 10^{-10}$ photons cm$^{-2}$ s$^{-1}$,  which is the expected point source flux sensitivity of GLAST for energies above
$\sim$ 1 GeV~\cite{deangelis_00}. This flux sensitivity  corresponds to a 5-$\sigma$ detection and 1 year of on-target observation.  
If GLAST achieves its expected sensitivity, 
then it  will be able to detect the neutralino signal for a significant 
portion of the parameter space. This is particularly true in light of the recently 
derived limit $m_{\chi}  < 500$ GeV~\cite{ellis_etal03} 
(vertical arrow in Fig.\ \ref{fig:susy}). 
If no annihilation signal is detected, the corresponding part of the parameter
space can be ruled out.  
For an energy threshold equal to 100 MeV we take $A_{eff}=10^4$ cm$^2$, 1 year of on-target
observation and $\Delta \Omega_{LMC} \simeq 1.13 \times 10^{-2}$ sr for the solid angle 
relevant for the collection of noise (since the LMC will be resolved by GLAST).
With the
sensitivity for this energy threshold  only a small 
portion of the parameter space will be detectable (dot-dashed line in Fig.\ \ref{fig:susy}).  
Clearly, 
the best chance for detection is by using moderate energy thresholds, such as the 1 GeV case, 
for which the backgrounds are relatively low but the source photons are still numerous.
To identify neutralino annihilation as the origin of the observed flux,
the spectrum and its characteristic features, such as
the cutoff at $E=m_{\chi}$, may be useful. 
In addition, monochromatic lines produced by neutralino annihilation 
(e.g., the $\gamma \gamma$ line at $E=m_{\chi}$) can be useful observational signatures, 
even though these lines are strongly suppressed (see, e.g.,~\cite{tasitsiomi_olinto02}).

The detectability prospects for existing and upcoming ACTs are less
optimistic.  The commonly assumed 
specifications are  $A_{eff}=10^{8}$ cm$^2$, $E_{th}=50$ GeV and 100 
hours of observation. For these specifications no models will be detectable.  
A large integration time ($\sim 1$ month) and effective area 
($\sim10^9$ cm$^2$)  (dotted line in Figure \ref{fig:susy}) would improve the chances for detection. 
However, such integration times are fairly long for ACT observations and such large effective areas for an energy threshold  
of $\sim$ 50 GeV are beyond the goals of existing and upcoming ACTs. Effective areas of order 
$10^9$ cm$^2$ are expected to be achieved by ACTs for energies $\sim 1$ TeV, but the number 
of $\gamma$-rays  produced by dark matter annihilation with energies 
$>$ 1 TeV is expected to be zero for models where the upper limit on $m_{\chi} < 500$ GeV holds.
\nopagebreak
\section{The synchrotron emission}
Neutralino annihilation not only generates neutral pions, but also a comparable number of charged pions.
In the presence of magnetic fields,  
the electrons and positrons produced in this way generate synchrotron radiation. 
Using again the Hill spectrum for the quark fragmentation, and
 taking into account
 synchrotron losses and inverse Compton scattering (ICS)
we calculated the synchrotron flux  
$F_{syn}$ in Jy (1 Jy=$10^{-23}$ ergs cm$^{-2}$ s$^{-1}$ Hz$^{-1}$).
\begin{figure}
\includegraphics[height=.315\textheight]{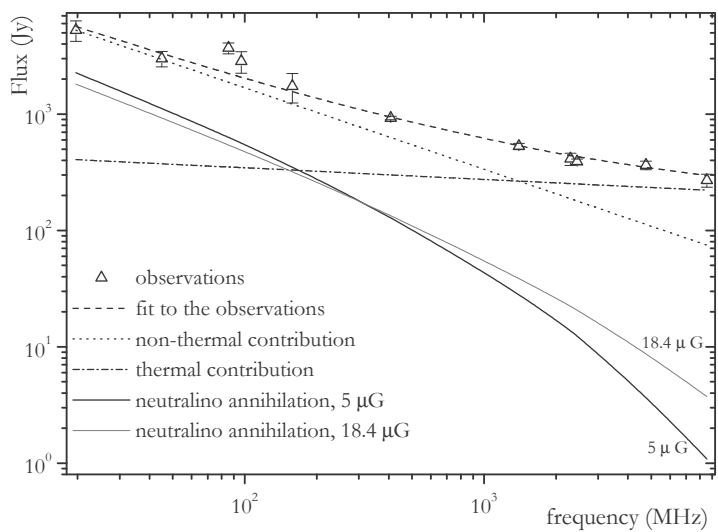}
\caption{The synchrotron flux in Jy as a function of frequency in the range from 19.7 to 8550 MHz. The open triangles
correspond to LMC data taken over the period from 1959 to 1991; the
dashed line is a best fit to the data assuming a decomposition into thermal and non-thermal emission. 
The dotted line corresponds to the non-thermal contribution, and the dot-dashed line to the thermal contribution. 
The flux from neutralino annihilation for $m_{\chi}=$ 50 GeV, $B_{\mu}= 5\ \mu G$, and
$\langle \sigma v \rangle \simeq 2 \times 10^{-26}$ cm$^3$ s$^{-1}$, is given by the thick solid line. 
The remaining line corresponds to $B_{\mu}= 18.4\ \mu G$, which is the maximum published estimate for
the total magnetic field. }
\label{results}
\end{figure}

In Figure \ref{results} I show $F_{syn}$ for $m_{\chi}=50$ GeV and  $\langle \sigma v \rangle= 2.2 \times 10^{-26}$  cm$^3$ s$^{-1}$, for 
the NFW dark matter profile. The thick solid line corresponds to $ B_{\mu}=5 \mu G$; 
the lighter line represents the results for the maximum total magnetic
field  estimated in literature (18.4 $\mu G$~\cite{chi_wolfendale93}). 
A compilation of radio observations of the LMC from 1959 to 1991 in the frequency range $(19.7- 8550)$ MHz 
is also shown \citep{haynes_etal91}. 
In the more realistic case of $B_{\mu}=5\ \mu G$, the neutralino induced synchrotron emission clearly may be part of the observed flux, 
but is lower than the total observed flux, especially at higher frequencies.
In the low frequency region, the signal increases enough to exceed the
thermal contribution to the observation. 

Clearly, as was the case for the $\gamma$-ray flux, a significant contribution
in the osberved radio-emission must be due to CRs.
The best hope for distinguishing the synchrotron emission generated by cosmic rays from that produced by neutralino annihilation
is to use the fact that the density profile of cosmic rays differs from the dark matter density profile.  While the dark matter halo extends significantly beyond the disk, the density of cosmic rays is expected to rapidly decrease at large radii.  If low radio frequency observations of the LMC are made with high angular resolution, the change from cosmic ray to neutralino dominance should be apparent as one moves away from the LMC disk. Currently there is no telescope that
could carry out observations of the LMC at frequencies less than about $20$ MHz.
Such low frequencies
 are extremely difficult to observe from the ground due to ionospheric
absorption and scattering. One promising ground-based 
project is the Low Frequency Array (LOFAR)~\cite{rottgering03} which should reach 
frequencies  down to $\sim$ 10 MHz and a flux sensitivity of a few $m$Jy  in 1 hour. 
However, the site of the instrument has not yet been 
decided, so the LMC may not necessarily be observable. 
To reach even lower frequencies, where ionospheric  absorption is very intense,
space-based instruments are required. The proposed Astronomical Low Frequency Array (ALFA)~\cite{jones_etal00} should reach down to $\sim 0.3$ MHz.
%%%%%%%%%%%%%%%%%%%%%%%%%%%%%%%%%%%%%%%%%%%%%%%%
%% BACKMATTER
%%%%%%%%%%%%%%%%%%%%%%%%%%%%%%%%%%%%%%%%%%%%%%%%

\begin{theacknowledgments}
This work was supported by NSF grant PHY-0114422 at the Kavli Institute for Cosmological Physics at the
University of Chicago.
\end{theacknowledgments}

%% If the bibliography is produced without BibTeX comment out the

\bibliographystyle{aipproc}   % if natbib is available

\bibliography{lmc_proceedings.bbl}

\IfFileExists{\jobname.bbl}{}
 {\typeout{}
  \typeout{******************************************}
  \typeout{** Please run "bibtex \jobname" to optain}
  \typeout{** the bibliography and then re-run LaTeX}
  \typeout{** twice to fix the references!}
  \typeout{******************************************}
  \typeout{}
 }

\end{document}